# Electronic structure of Gd pnictides

Paul Larson and Walter R. L. Lambrecht
*Department of Physics,*
*Case Western Reserve University, 10900 Euclid Ave*
*Cleveland, OH 44106*

A computational study of the electronic structure and magnetic properties of Gd-pnictides is reported. The calculations were performed using a full-potential linear muffin-tin orbital (LMTO) method within the so-called LSDA+U approach, which adds Hubbard-U correlation effects to specified narrow bands in a mean-field approach to the local spin density approximation (LSDA). Here both the Gd $4f$ and $5d$ states are subject to such corrections. The $U_f$ values were determined semi-empirically by using photo-emission and inverse photoemission data for GdP, GdAs, GdSb and GdBi. In contrast to $U_f$ which represents narrow band physics, $U_d$ represents a quasiparticle self-energy correction of the LSDA gap underestimate. The $U_d$ value was adjusted using optical absorption data for semiconducting GdN above its Curie temperature. Below the Curie temperature, however, in the ferromagnetic state, the gap becomes almost zero. The other Gd pnictides are found to have a small overlap of the conduction band at the $X$ point and the valence band at the $\Gamma$ point in the majority-spin channel. A small gap opens in the spin-minority channel of GdP and GdAs, which are thus half-metallic. This spin-minority gap closes in semimetallic GdSb and GdBi. While GdN is found to be ferromagnetic, the other Gd-pnictides are found to be antiferromagnetic, with ordering along [111]. From calculations with different magnetic configurations, a Heisenberg model with first and second nearest neighbor exchange parameters is extracted. The Heisenberg model is then used to predict Curie-Weiss and Néel temperatures and critical magnetic fields within mean field and compared with experimental data. The trends are found to be in good agreement with the experimental data.

## I. INTRODUCTION

The emergence of the field of spintronics[1] has led to a search for new magnetic semiconductors. These include primarily dilute magnetic semiconductors (DMS's), most often based on transition metal doping of traditional III-V or II-VI semiconductors[2]. However, some intrinsic magnetic semiconductors also exist based on the presence of rare-earth elements with open $4f$ shells. A well known example of this is EuO which has a moment of 7 $\mu_B$, a band gap of 1.12 eV at room temperature, and a Curie temperature $T_c$ = 69 K.[3] Below $T_c$ this material is known to undergo a redshift (the gap narrows).

Like EuO the Gd pnictides form in the rocksalt crystal structure. While experimental evidence shows that GdP, GdAs, GdSb, and GdBi are antiferromagnetic (AFM) and metallic[4], GdN has been reported to be ferromagnetic (FM)[5–7] and may be a semiconductor. While early reports ascribed the ferromagnetism to the Ruderman-Kittel-Kasuya-Yoshida (RKKY) carrier mediated mechanism due to impurity (oxygen) or defect (nitrogen vacancies) related doping, recent studies[7, 8], claim that even purely stoichiometric GdN is ferromagnetic with a critical temperature of $T_c$=58 K. It has been suggested that the mechanism for strong ferromagnetism in GdN is different from that in EuO[9], but how these differences will affect their performance as magnetic semiconductors is unknown. A first-principles understanding of the nature of the magnetic interactions is also lacking. EuS and EuSe are also FM while EuTe is AFM[3], so the fact that the other Gd pnictides are AFM and metallic[4] needs to be explored further in order to understand the magnetic and electronic properties of GdN.

It is also still under dispute if GdN is a semimetal with a small band overlap, a zero gap semiconductor, or a finite gap semiconductor. The situation is similar to that of its transition metal cousin, ScN, which was long thought to be a semimetal because of the carrier concentrations of order $10^{20}/\text{cm}^3$ which were typically observed.[10] It was only when activated nitrogen sources were developed for III-N growth that it became possible to reduce the carrier concentration in ScN to $10^{17}$ /cm$^3$ levels[11, 12] and thereby to demonstrate that ScN is in fact a semiconductor. Optical absorption studies show a direct gap of about 2 eV and only recently, evidence for a smaller indirect gap of about 1 eV became available.[13] Band structure calculations in the local density approximation (LDA) give a very small overlap for ScN and GdN lending further credibility to the notion that these materials might be semimetallic. However, LDA is well known to underestimate band gaps. Recent calculations using exact exchange (EEX) [14] or screened exchange (SX) [15] as well as arguments based on the GW approach[16] all agree that ScN is an indirect gap semiconductor. A similar situation occurs in GdN but it is further complicated by the presence of the half-filled $4f$ shell, which induces a spin-splitting of the conduction and valence bands. The carrier con-centration has been reduced from earlier measurements but is still of order $1.9 \times 10^{21}$ cm$^{-3}$[6]. Direct resistivity measurements[17] as well as the measurement of an optical absorption edge of about 1 eV[6] indicate that GdN has a band gap at least at room temperature. Recent measurements[8] however find an insulator to metal transition going through the critical temperature ($T_c$), so the presence or absence of a band gap appears to depend on the magnetic configuration.

## II. COMPUTATIONAL METHOD

Electronic structure calculations were carried out within the LSDA+U approach. This method is based on the local spin density approximation (LSDA) to density functional theory (DFT)[18] using the exchange-correlation parameterization of von Barth and Hedin[19] but is complemented with Hubbard-$U$ corrections treated in a mean-field approximation. Our calculations use a full-potential linearized muffin-tin orbital (FP-LMTO) approach introduced by Methfessel and van Schilfgaarde.[20] This method uses an optimized basis set consisting of muffin-tin orbitals with smoothed Hankel functions as envelope functions. The smoothing radii and $\kappa$ (Hankel function decay parameter) values were carefully adjusted to optimize an efficient basis set with one $s$, $p$, and $d$ function on the pnictogen site, two $s$ and $p$ and a single $d$ and $f$ function on each Gd site. The smooth interstitial quantities are calculated using a fast Fourier transform mesh and the Brillouin zone integrations were carried out using the tetrahedron method[21, 22] with a well-converged $k$-mesh based on a division of the reciprocal unit cell in $6 \times 6 \times 6$ divisions.

The partially filled and strongly correlated localized $f$ orbitals were treated using the LSDA+U formalism,[23–26] where the double counting terms are subtracted within the fully-localized limit (FLL) which best describes the localized nature of the $4f$ orbitals. The present implementation in the van Schilfgaarde *lmf* program follows the rotationally invariant formulation of Liechtenstein et al.[25] and includes non-spherical terms, described in terms of the Slater-Coulomb integrals $F^k$, with $k = 0, 2, 4, 6$. As is customary, it is assumed that only $U = F^0$ is strongly screened, whereas $F^2$–$F^6$ behave as in the free atom. Furthermore, the ratios of $F^6/F^2$ and $F^4/F^2$ are well known to be almost independent of the element and thus these 3 parameters can be reduced to one effective $J$ parameter given by

$$J = (286F^2 + 195F^4 + 250F^6)/6435 \qquad (1)$$

In the Gd pnictides the $4f$-shell is half-filled and thus spherically symmetric, so the $J$ parameter does not lead to a splitting of the $f$-states but just enters in the combination $U-J$.[26] We symmetrize the density matrix $\rho_{mm'}$ and the associated potential $V_{mm'}$ according to the space group operations of the crystal structure and allow them to become self-consistent due to the partial hybridization of the $4f$ states with the other states in the system. While both the around mean field (AMF)[23] and fully localized limit (FLL)[25] versions of LSDA+U and the Petukhov-Mazin scheme for mixing AMF and FLL[27] are included in the program, FLL was used here for all the calculations because of the highly localized nature of the $4f$ orbitals.

We did not attempt here to calculate $U_f$ and $J_f$ values independently from constrained LSDA calculations[24] but rather used a semi-empirical approach. We start from the values of the integrals $F^k$, for $k$=0,2,4,6 obtained from Hartree-Fock calculations of the elements[28]. We scaled $F^0$ to the $U$ parameter so that the splitting of the occupied and empty $4f$ bands, agrees reasonably well for GdP, GdAs, GdSb and GdBi for which photoemission and inverse photoemission data are available.[29] We note that the splitting between occupied and unoccupied $f$ states in these compounds appears to be nearly independent of the pnictogen, so a single $U_f$ is adopted for all materials and also used for GdN. For $J_f$ we used Eq.(1), using unscreened atomic values for the $F^k$ integrals as tabulated by Mann[28]. Here no screening is performed since the $J$ represents the internal splittings by quantum number $m$ for each $l$ which are essentially atomic splittings unaffected by the solid environment. The values of $U_f$ and $J_f$ used in the calculation here are $U_f$ = 8.0 eV and $J_f$ = 1.2 eV.

As was already qualitatively discussed in the introduction, and will become clearer in the next section, the Gd $5d$ bands form the conduction band and are thus essentially empty. In fact, without any further correction, a small band overlap would occur between the lowest conduction band at $X$ and the highest valence bands at $\Gamma$ in GdN. We can use the same LSDA+U approach to shift the $d$ bands. Although the physics for this shift is different in origin, it can easily be implemented in the same way. The main reason for the need for a $d$-band shift is the usual underestimate of band gaps by LSDA. Starting from the quasiparticle GW theory perspective, one needs to evaluate primarily the statically screened exchange term. As was shown by Maksimov and Mazin,[30] the latter is the most "non-local" term because it contains the long-range exchange decaying like $1/r$ in a semiconductor. Even if we start out from a border line semimetal, the screening with a typical carrier concentration of a semimetal is still long-range as argued in Ref. 16. Ultimately, however, we just need a matrix element of this between conduction band states, which here are almost purely Gd $5d$ like. Thus it comes down to a shift of the Gd $5d$ states. Since the LSDA+U potential is of the form,

$$V_{mm'} = -(U - J)(\rho_{mm'} - \frac{1}{2}\delta_{mm'}) \qquad (2)$$

in terms of the density matrix $\rho_{mm'}$ within the $d$-band manifold, it reduces to an upward shift by $(U - J)/2$ for empty orbitals. Thus, we can add a shift of the conduction band by adding a $U_d$. For simplicity we set $J_d$=0.

We pick the $U_d$ value in GdN so as to adjust the lowest optical direct transition at $X$ to the optical gap above the Curie temperature as will be explained in section III A. The value found in this manner is $U_d = 3.4$ eV. The same $U_d$ value is also used from the other Gd-group-V compounds, but this may be a slight overestimate. One might expect that in the semimetallic state the screening could be slightly stronger. Nonetheless, as we will see in the results section all of these materials stay semimetallic.

## III. RESULTS

Gd pnictide compounds form in the rocksalt crystal structure. The equilibrium lattice constants obtained by energy minimization agree well with experimental values as seen in Table I. We also allowed for a possible tetragonal distortion but found that all of these materials stay cubic.

TABLE I: Experimental[5, 29] and calculated equilibrium lattice constants for Gd-pnictides.

| (Å) | Exp. | Theor. |
|---|---|---|
| GdN | 4.98 | 4.98 |
| GdP | 5.71 | 5.65 |
| GdAs | 5.86 | 5.78 |
| GdSb | 6.22 | 6.09 |
| GdBi | 6.30 | 6.36 |

The electronic and magnetic properties of the Gd pnictides involve several aspects, including the band gap, the position of the $4f$ states, and the magnetic exchange parameters which describe the magnetic properties of the system. The results will be organized as follows. First, we will discuss the question of the band gap in GdN. Next, we will present the band structures of the other pnictides. Finally, we will present our results on the magnetic energy configurations and magnetic exchange parameters for this class of materials.

### A. Band gap in GdN

We start our discussion with an overview of the band structure of GdN obtained in both LSDA and LSDA+U as shown in Fig. 1. In LSDA, the occupied $4f$ bands cut through the valence band at about -3.4 eV below the Fermi level and the empty states lie about 4 eV above the Fermi level. The band structure is semimetallic. In the LSDA+U calculation, the $f$ bands are shifted farther away from the Fermi level. They form a narrow band of majority spin at about 6.7 eV below the Fermi level, well separated from the mainly N $p$ like valence bands, and a somewhat wider set of bands of minority spin at 7–8 eV above the Fermi level. The position of the occupied bands is in good agreement with the experimental photoemission data which place these bands at $-7.8$ eV.[8, 31] If we use a higher $U_f$ value like 9-10 eV, one can push the occupied levels down a bit further but then the unoccupied ones also move to higher energies. So, the $f$ bands do not exactly move by $(U - J)/2$ from the Fermi level due to hybridization effects with other bands.

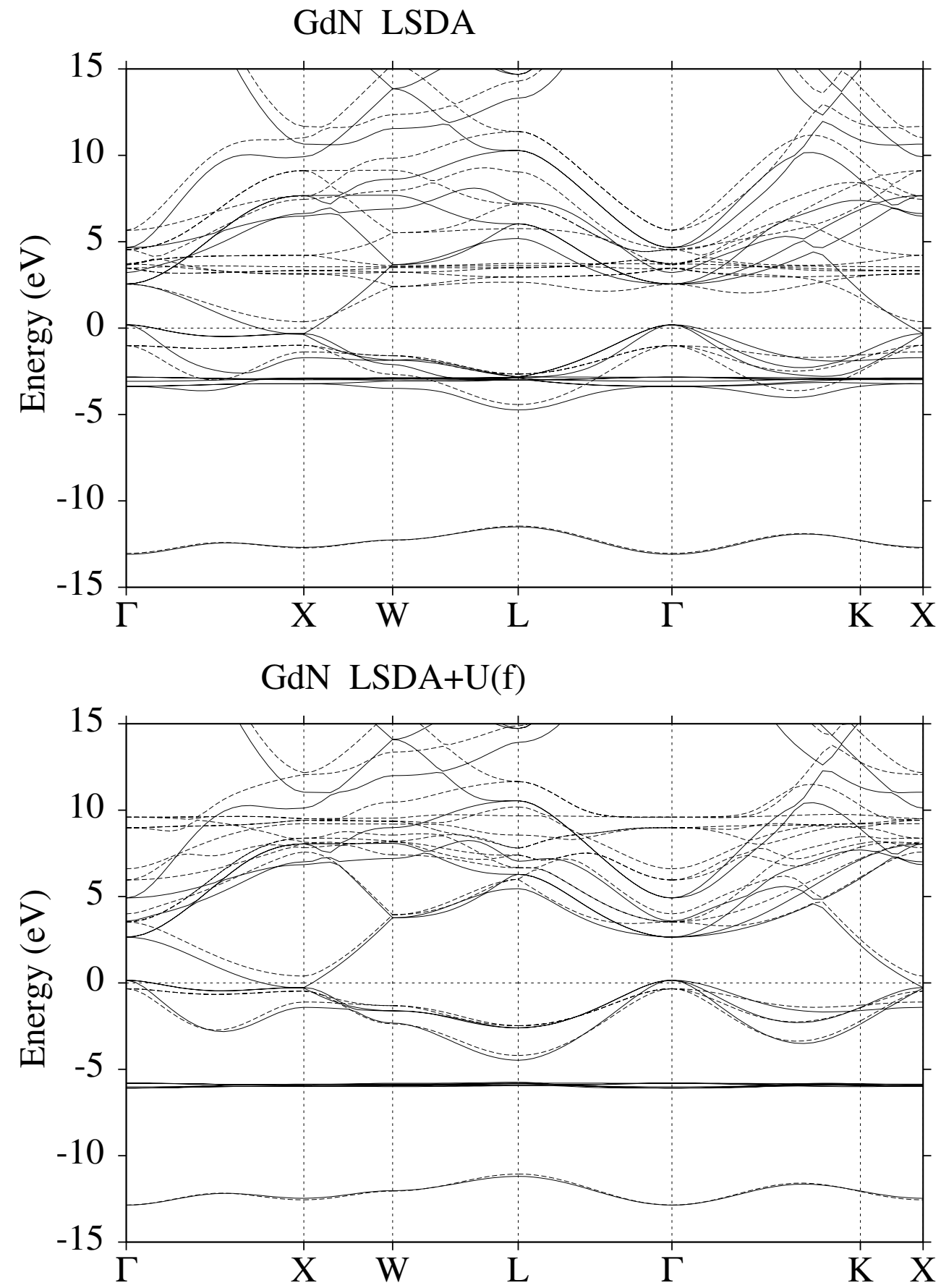

FIG. 1: Electronic structure of GdN for (a) LSDA and (b) LSDA+U with $U_f = 8.0$ eV and $J_f = 1.2$ eV. Solid lines: majority spin, dashed lines: minority spin

If we include the $U_f$ Hubbard-U terms but not the $d$-band shift, the Gd $d$ band still dips slightly below the Fermi level at the $X$ point. This is shown in more detail in Fig.2b. A large spin-splitting of these bands arises due to the exchange splitting introduced by the localized magnetic moments of the $4f$ states. We note that the spin-splitting is inverted between the valence band and the conduction band. This results from the fact that the N $2p$ bands of majority spin are being pushed up by their interaction with the lower Gd $4f$ majority spin band. The Gd $5d$ states however are orthogonal to the $4f$ and thus simply feel the different exchange potential, lowering the energy for majority spin electrons. This leads to a band gap for minority spin but a slight band overlap for the majority spin. In fact, the $d$-like conduction band crosses the N $p$-like valence band of majority spin just before the $X$ point.

Experimentally, it is known that GdN at room tem-perature has a direct onset of optical absorption at 0.98 eV.[3]. Since this temperature is clearly above the Curie temperature, it corresponds to the paramagnetic state. This means that the spins are randomly oriented and no net spin polarization should be induced in the conduction or valence bands. One can invoke a "virtual crystal approximation" which means this situation can approximately be described by averaging the slightly negative majority spin gap and about 0.9 eV minority spin gap at $X$. This approach has been used to explain Eu-rich EuO.[32] In fact, the lowest direct optical transition occurs at $X$ as can be seen from the overview band structure plots. This would lead to a spin averaged gap of only 0.42 eV. By adding a $U_d = 3.4$ eV, the empty $d$ bands shift up and now a gap appears for both majority and minority spin as can be seen in Fig. 2b. In fact, with this value of $U_d$, the majority spin direct gap at $X$ is 0.57 eV and the corresponding minority spin gap is 1.39 eV, leading to an average gap of 0.98 eV, fit to the experimental value. However, this means that the indirect gaps between $\Gamma$ and $X$ are now only 0.12 and 1.25 eV for majority and minority spin. Again, in the paramagnetic state, this predicts an indirect gap of about 0.685 eV. The latter has not yet been detected experimentally, but this is no surprise since an indirect absorption edge is difficult to unambiguously identify in the presence of a defect induced band gap tail. Including spin-orbit coupling further reduces the gaps to 0.09 eV for the smallest indirect $\Gamma - X$ gap and 0.50 eV for the $X$-point direct gap for majority spin. Thus the calculations predict that in the ferromagnetic state, the system is very close to a metal-insulator transition. Given that there could easily still be an uncertainty of about 0.1 eV on these gaps, this agrees well with the observation of a metal-insulator transition at the Curie temperature as reported by Leuenberger et al.[8] We furthermore observe that even if the one-electron gap is slightly positive, electron-hole coupling may lead to the spontaneous formation of a metal-insulator transition by exciton condensation into a correlated electron-hole liquid.[33] The results for the gaps of GdN are summarized in Table II. We also note that even in the paramagnetic state, it would be possible to align the magnetic moments in an external magnetic field. This should then similarly lead to a closing of the band gap.

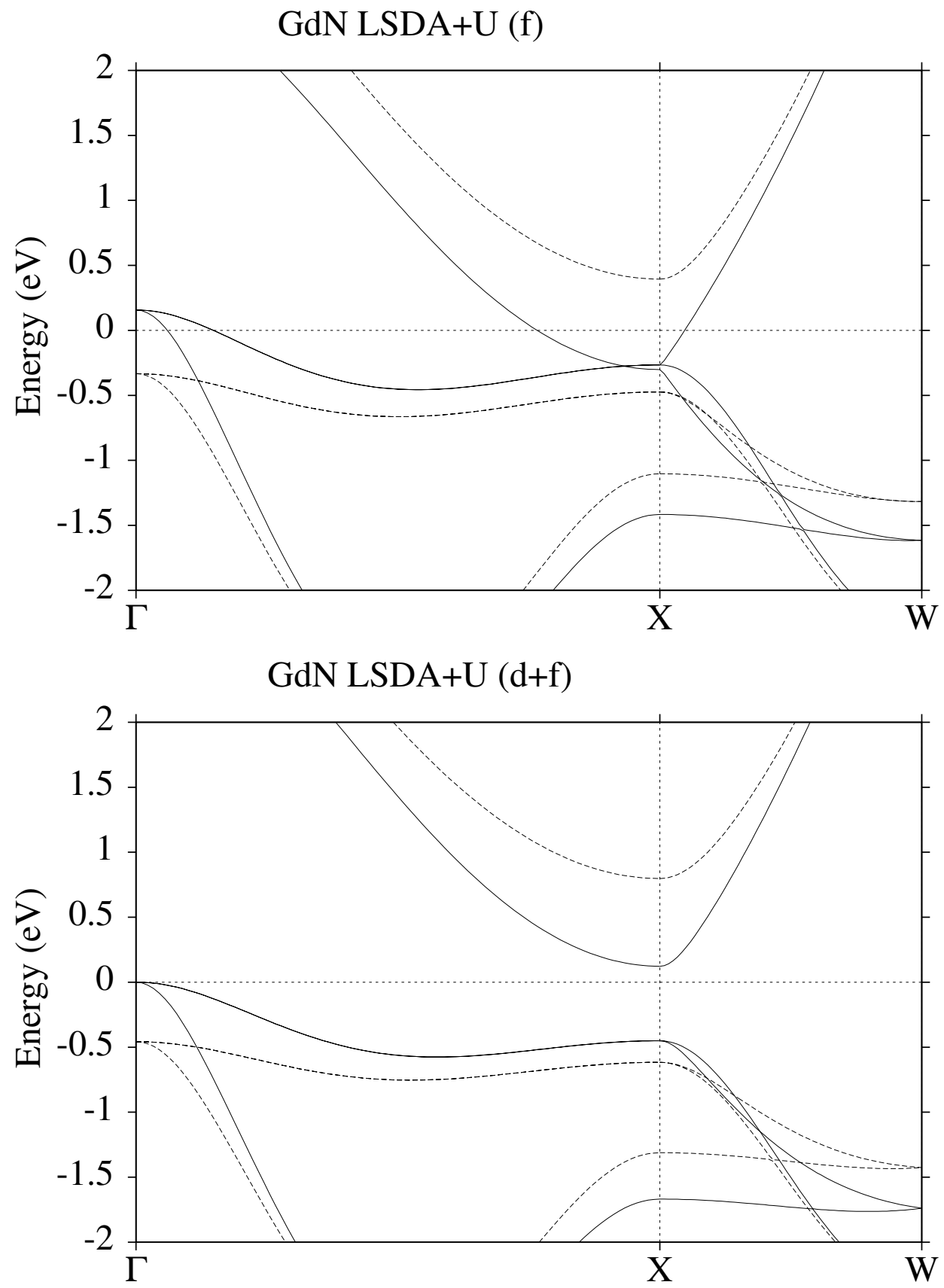

FIG. 2: Electronic structure of GdN for (a) LSDA+U of $U_f$ = 8.0 eV and $J_f$ = 1.2 eV on the Gd $4f$ orbital correction and (b) with an addition correction of $U_d$ = 3.4 eV and $J_d$ = 0 eV on the Gd $5d$ orbitals. Solid lines: majority spin, dashed lines minority spin.

TABLE II: Band gaps (in eV) in GdN in various approximations.

| Approximation | gap | ↓↓ | ↑↑ | average |
|---|---|---|---|---|
| $U_f$ no $U_d$ | direct at $X$ | −0.001 | 0.85 | 0.42 |
| | indirect $\Gamma - X$ | −0.42 | 0.73 | 0.16 |
| $U_f$ and $U_d$ | direct at $X$ | 0.57 | 1.39 | 0.98 |
| | indirect $\Gamma - X$ | 0.12 | 1.25 | 0.68 |
| $U_f$ and $U_d$ | direct at $X$ | 0.50 | 1.41 | 0.96 |
| + SO-coupling | indirect $\Gamma - X$ | 0.09 | 0.82 | 0.46 |

Finally, we comment on some other recent papers on this topic. In the earlier work by Petukhov et al.[34] the $4f$ states were treated as partially filled core states. In this approach, however, the interaction between Gd $4f$ states and N $2p$ states is not included and the spin-reversal of the valence band and conduction band states is not obtained. In that case, the gap for spin up and spin down is about the same and the metal-insulator transition would not be obtained. In the previous non-self-consistent LSDA+U calculation by Lambrecht[16] the gap was predicted to be slightly larger than obtained here even in the ferromagnetic state, but the redshift of the gap also plays an important role. Finally, Duan et al.[35] also used LSDA+U calculations for the $f$ states but did not include a gap shift for the $d$ bands. They obtained a semimetal situation in agreement with our results of Fig.2a. They observed that under lattice constant expansion, a gap would open. This is easily explained, since with a larger lattice spacing, the dispersion of the Gd $d$ band will be reduced. We also note that our calculated gap agrees well with a recent GW calculation by van Schilfgaarde et al.[36] For completeness, we also mention here the paper by Aerts et al.[37] which used a self-interaction correction (SIC) approach. This paper however does not discuss the band structure details or existence of a gap. A somewhat deeper position of the occupied $f$ bands at -13 eV is obtained in that work as well as a lower position of the unoccupied $f$ bands at about 2.5 eV above the Fermi level. Our present results, which are in close agreement with the other LSDA+U calculations, appear to be in better agreement with the X-ray absorption data of Leuenberger et al.[8] which show 3 peaks between 5 and 10 eV above the onset in the N $K$-edge spectrum, and a strong dichroic signal. Further work will be necessary to analyze these spectra in detail. This could be due to the hybridization of the minority spin Gd $4f$ states at about 7 eV above the VBM with N $2p$ states. Unfortunately, there do not appear to be any inverse photoemission experiments on GdN which would offer a more direct determination of the position of the empty $4f$ states. Finally, a totally different approach to the correlation and cohesive energy was introduced by Kalvoda et al.[38] but does not include results on the band structure.

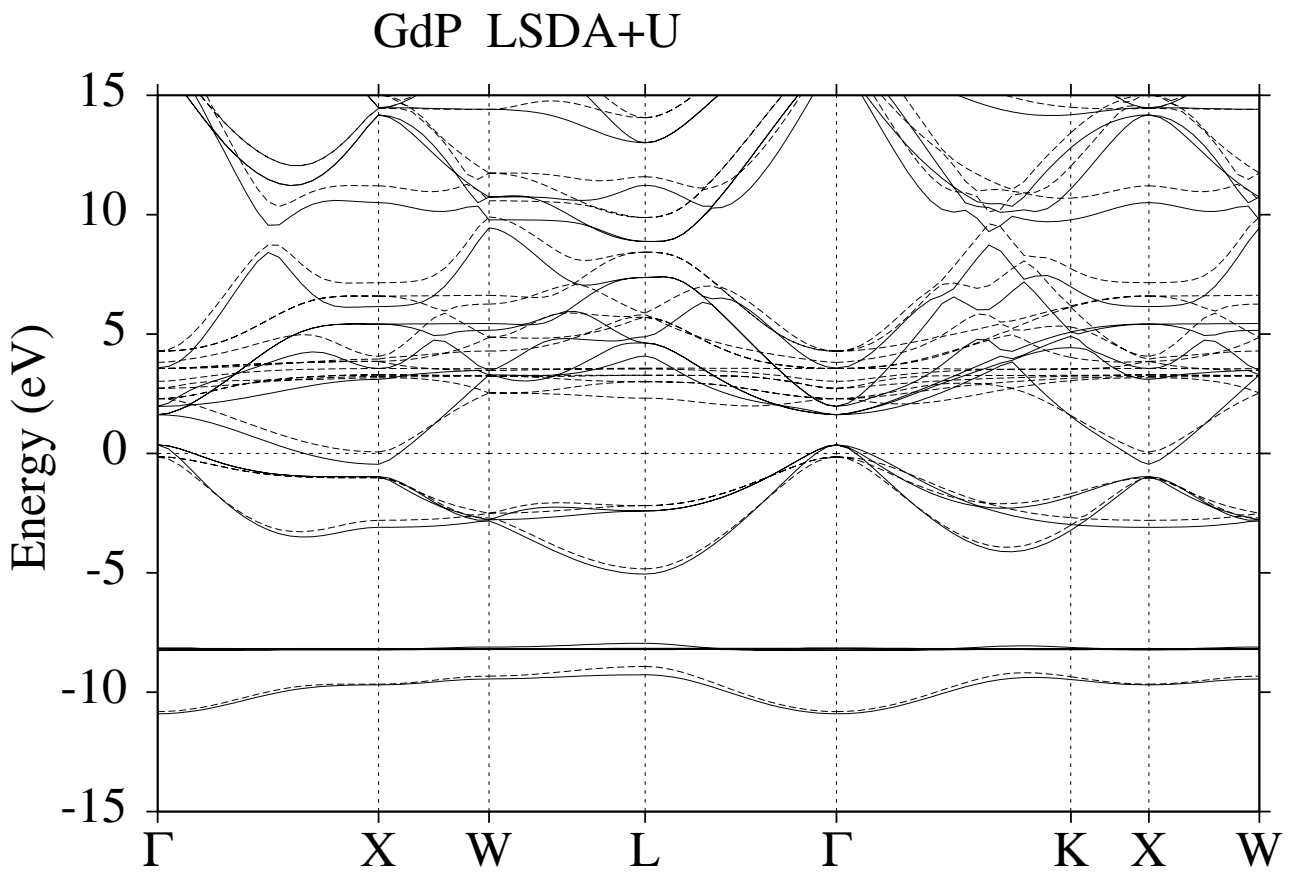

FIG. 3: Electronic structure of GdP for LDA+U correction of $U_f = 8.0$ eV and $J_f = 1.2$ eV on the Gd $4f$ orbitals and additional LSDA+U correction of $U_d = 3.4$ eV and $J_d = 0$ eV on the Gd $5d$ orbitals. GdP is predicted to be half-metallic with a spin-minority indirect gap of 0.17 eV.

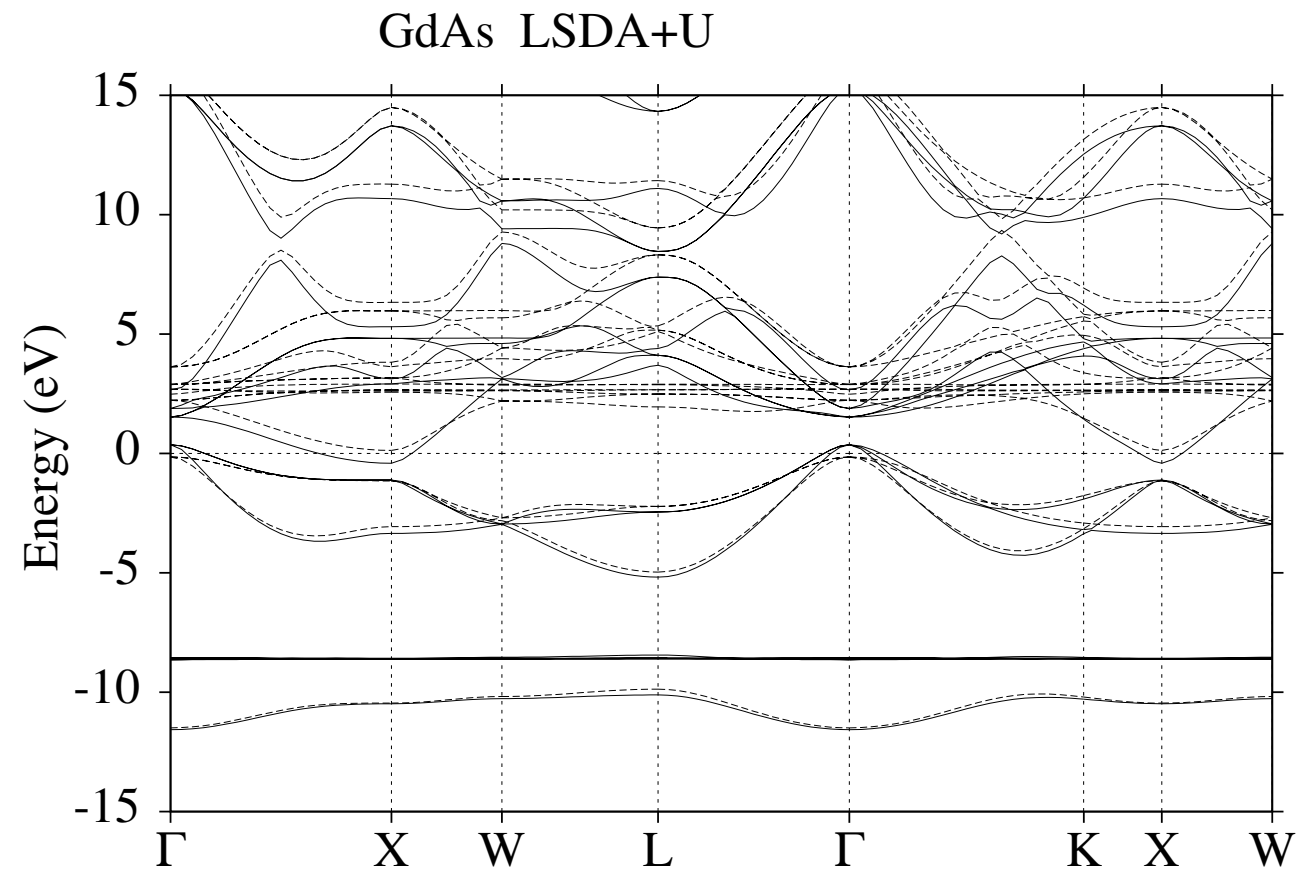

FIG. 4: Electronic structure of GdAs for LDA+U correction of $U_f = 8.0$ eV and $J_f = 1.2$ eV on the Gd $4f$ orbitals and additional LSDA+U correction of $U_d = 3.4$ eV and $J_d = 0$ eV on the Gd $5d$ orbitals. GdAs is predicted to be half-metallic with a spin-minority gap of 0.16 eV.

### B. GdP, GdAs, GdSb, and GdBi

The band structures of GdP, GdAs, GdSb, and GdBi, shown in Figs. 3, 4, 5, 6 respectively, are very similar to each other, but are different in several ways from that of GdN. In GdN (Fig. 1b), the Gd $4f$ orbitals lie fairly symmetrically about the Fermi level ($E_F$) at about +7 eV and −7 eV. The unfilled bands are fairly broad and hybridize with the conduction band states. In GdP, GdAs, GdSb, and GdBi the empty Gd $4f$ states move closer to $E_F$, starting around +4.5 eV in GdP and moving to +2.0 eV in GdBi, and become narrower. The empty Gd $4f$ states move more into the low density of states region near the bottom of the conduction band.

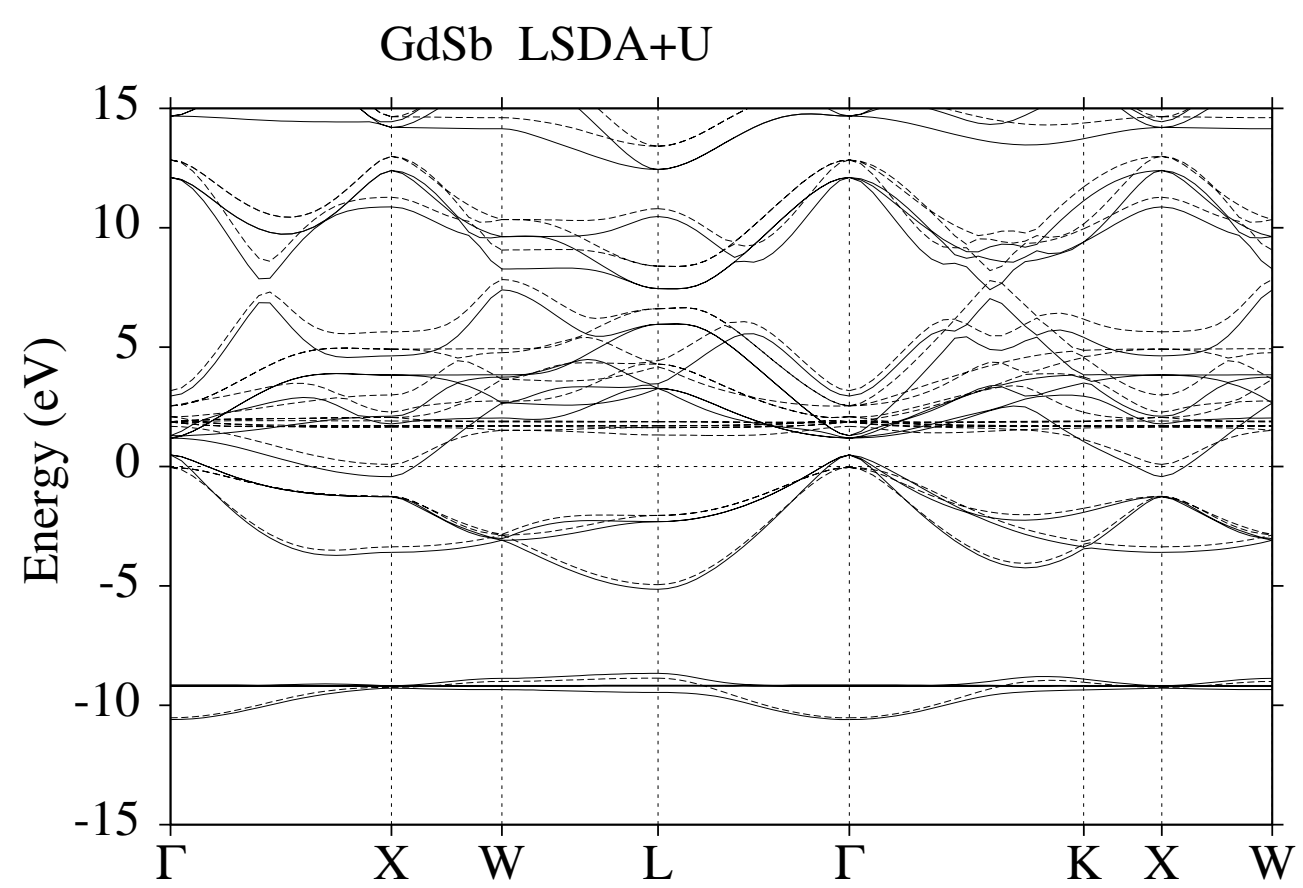

FIG. 5: Electronic structure of GdSb for LDA+U correction of $U_f = 8.0$ eV and $J_f = 1.2$ eV on the Gd $4f$ orbitals and additional LSDA+U correction of $U_f = 3.4$ eV and $J = 0$ eV on the Gd $5d$ orbitals.

The filled $4f$ states also move down in energy by about 1.5 eV, so the splitting is not exactly the same as in GdN. Photoemission measurements[29] find that both the filled and unfilled Gd $4f$ bands move down in energy going from P-Bi, especially from Sb to Bi, in qualitative agreement with our calculations. In Figure 7 we show the average position of the occupied and empty $4f$ states relative to the Fermi level across the series, compared with the photoemission data and inverse photoemission data. For GdN, we have only photoemission data available[8], no inverse photoemission data. However, some features in the X-ray circular magnetic dichroism (XMCD) and X-ray absorption at about 5-8 eV above the conduction band edge are probably related to the empty $4f$-levels[8]. Our results indicate a more linear behavior from P to Bi, in contrast to the photoemission data[29] which indicate only a small change from P to Sb and a more sudden change from Sb to Bi. Generally speaking if we adjust $U_f$ to reproduce the occupied $f$-levels, the unoccupied ones lie lower in our LSDA+U calculation than in experiment. Similarly, adjusting $U_f$ to get agreement with position of the unoccupied $4f$ states pushes occupied $4f$ states considerably below where they are seen in experiment. The origin of this discrepancy is not clear. Recent GW calculations,[36] for ErAs find a higher position for the empty $4f$ states than LSDA+U, indicating that energy dependent self-energy corrections are required beyond LSDA+U.

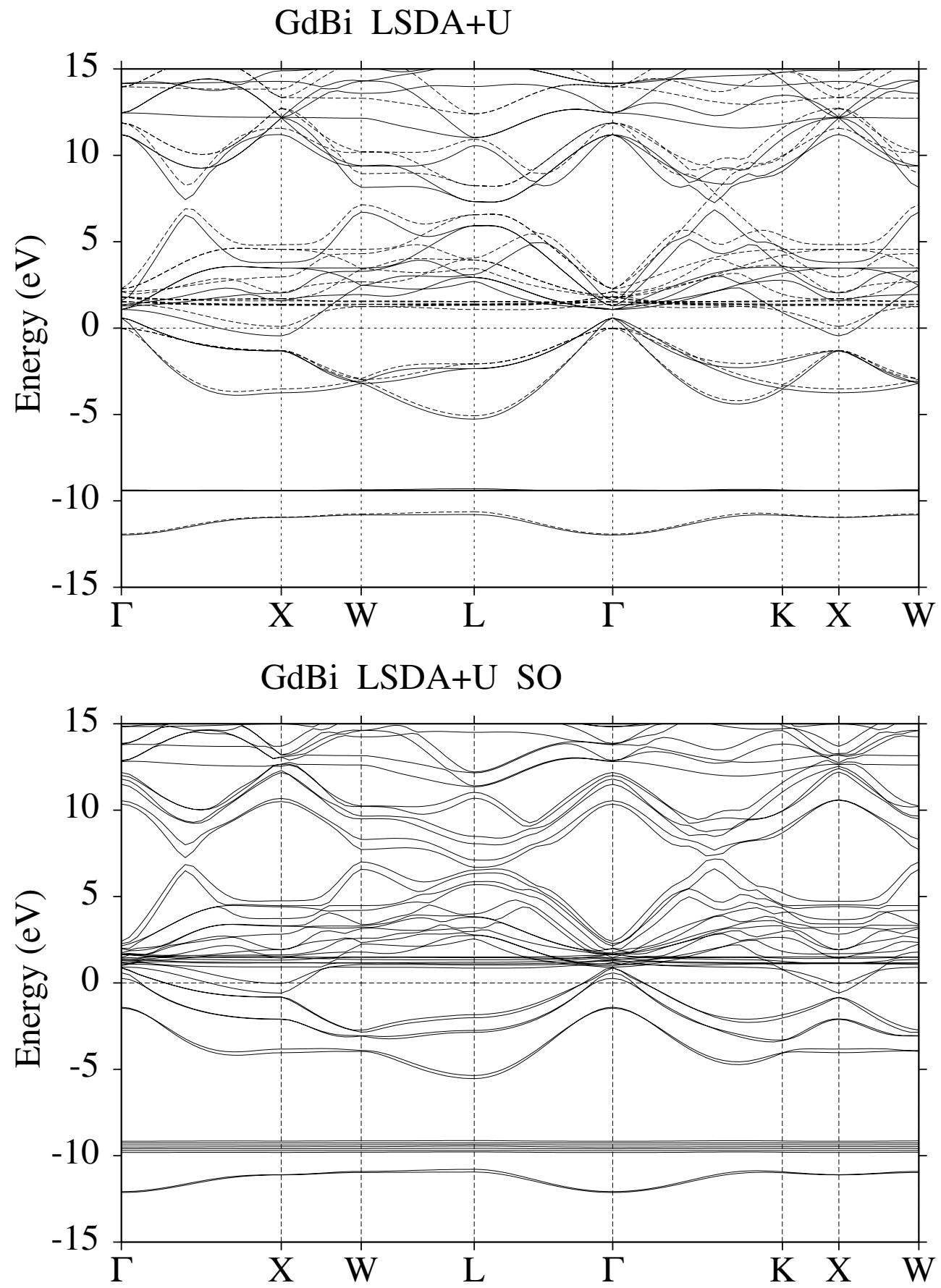

FIG. 6: Electronic structure of GdBi for LDA+U correction of $U = 8.0$ eV and $J = 1.2$ eV on the Gd $4f$ orbitals and additional LDA+U correction of $U = 3.4$ eV and $J = 0$ eV on the Gd $5d$ orbitals. GdBi (a) without spin-orbit (SO) and (b) with spin-orbit (SO) which is large for Bi 6p.

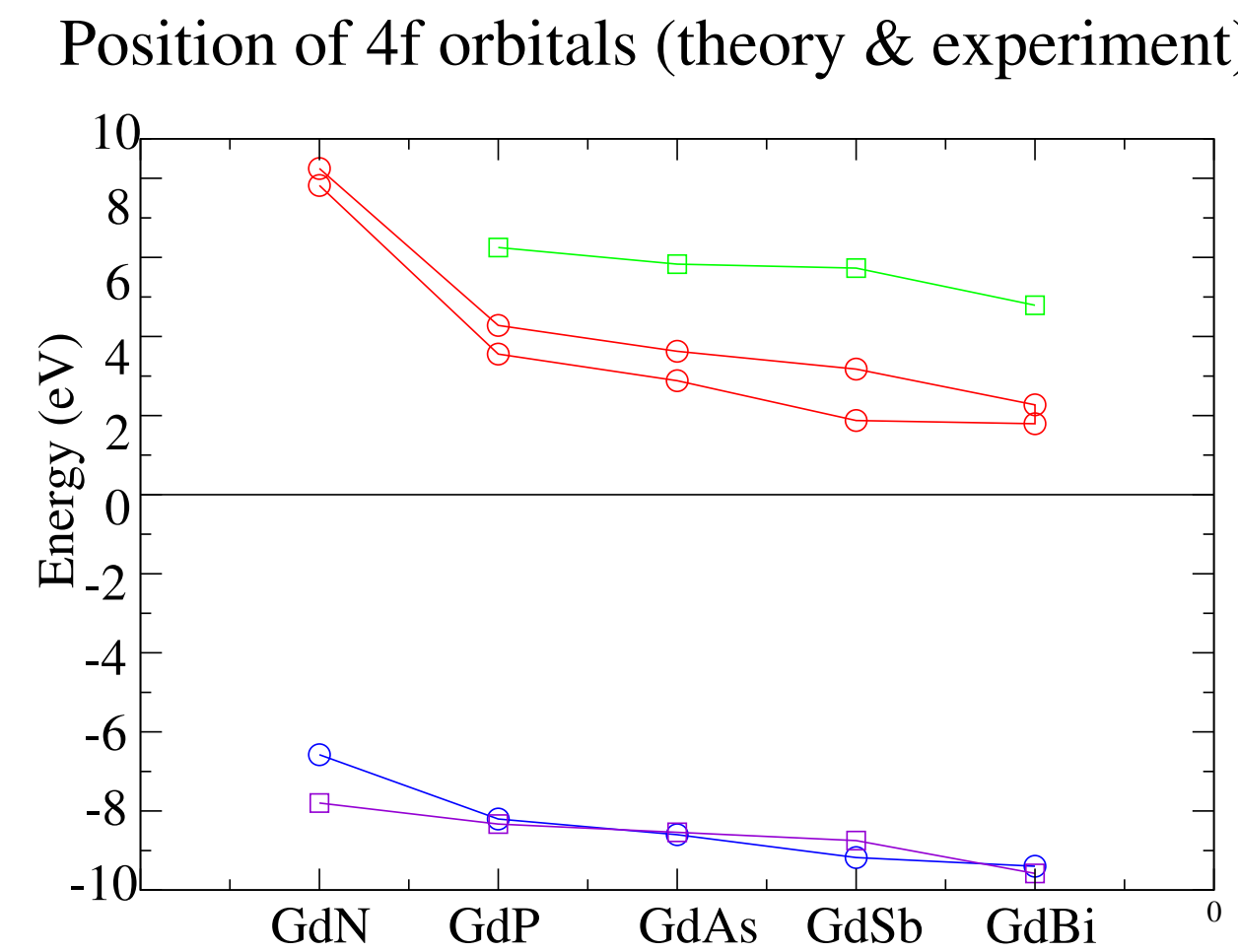

FIG. 7: Average position of occupied and unoccupied $4f$ states relative to the Fermi level in the Gd-pnictides for the calculation (circles) and experiment (squares[8, 29]). The two sets of values for the unoccupied $4f$ states indicates the upper and lower limits of this fairly wide band.

The overlap of the Gd $5d$ bands with the valence band anion $p$ states is larger in GdX with X=P,As,Sb,Bi than GdN which should increase the screening and the $U_d$ parameter used. However, even when we apply the same $U_d$ shift as in GdN, these materials stay semimetallic at least when we consider the majority spin gap, since the overlap is larger to begin with in the LSDA. The minority spin states, however, maintain a gap for GdP and GdAs but not for GdSb and GdBi. The gap first of all decreases along this series from 0.2 eV in GdP to 0.1 eV in GdBi and become further reduced by including spin orbit coupling. These values are given in Table III. Therefore, GdP and GdAs in principle are half-metallic in their ferromagnetic state. However, as will be shown in the next section, they actually prefer an antiferromagnetic alignment. Thus this half-metallic behavior can probably only be observed in an external saturating magnetic field, starting from the paramagnetic state. If we consider the spin averaged indirect gaps, GdP and GdAs appear to still be very narrow gap semiconductors. In GdSb and GdBi, the trend continues but now even the minority gaps become negative once spin-orbit coupling is included. So, these are no longer half-metals. As noted earlier our $U_d$ shift is probably somewhat of an overestimate. With a stronger screening and smaller $U_d$ the spin-minority gap in GdP and GdAs will reduce further. This agrees with experimental reports of the semimetallic behavior of all these compounds.[4, 7] However, for a more quantitative determination of the necessary $U_d$ shifts, we would need experimental data on the carrier concentrations and hence size of the Fermi surface pockets as was for example carried out in Ref. 39. For GdBi, we explicitly show the bands with and without spin-orbit coupling in Figure 6, because one may expect these effects to be strongest in the material with the heaviest pnictogen element, Bi. Even in this case, the general features of the band structure as discussed above are maintained but the reduction of the minority spin gap by spin orbit coupling is indeed more significant. It turns into a band overlap of 0.29 eV.

TABLE III: Band gaps and overlaps (in eV) in Gd-pnictides other than GdN with spin orbit included. The values in parentheses are without spin orbit coupling.

| (eV) | majority spin | minority spin | average |
|---|---|---|---|
| GdP | -0.87 (-0.81 ) | 0.17 (0.20) | -0.35 (-0.30 ) |
| GdAs | -0.71 (-0.78 ) | 0.16 (0.27) | -0.27 (-0.52) |
| GdSb | -1.25 (-0.91) | -0.01 (0.13) | -0.63 (-0.39) |
| GdBi | -1.17 (-1.04) | -0.29 (0.10) | -0.73 (-0.47) |

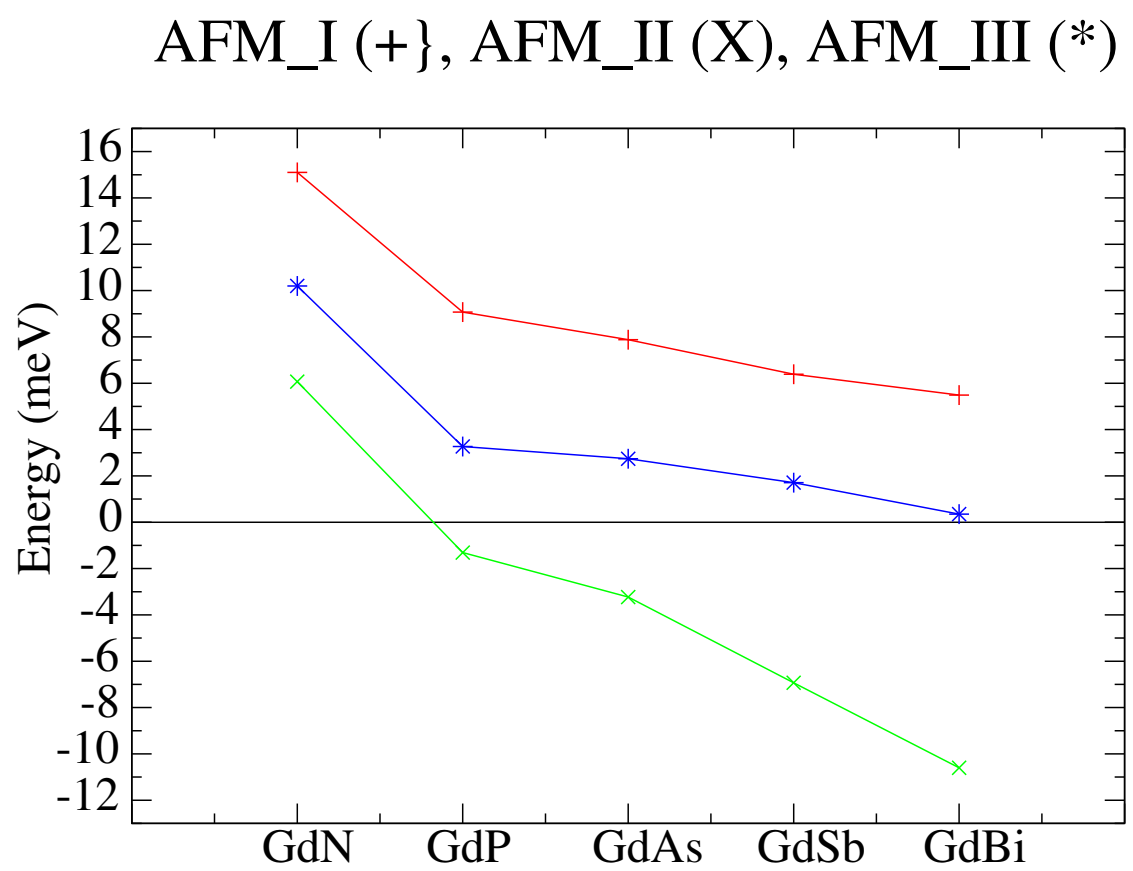

FIG. 8: Energy of $AFM_I$, $AFM_{II}$, and $AFM_{III}$ states with respect to the FM state. Only in GdN does FM has the lowest energy while for the others the $AFM_{II}$ state has lowest energy.

### C. Magnetic exchange parameters

The magnetic properties of Gd pnictides are puzzling, especially the change from ferromagnetic (FM) ordering in GdN to antiferromagnetic ($AFM_{II}$) ordering in GdP, GdAs, GdSb, and GdBi[4]. The simplest FM and AFM configurations have been described as FM, $AFM_I$, $AFM_{II}$, and $AFM_{III}$[40], where $AFM_I$ describes alternating spins in the [001] direction, $AFM_{II}$ describes alternating spins in the [111] direction, and $AFM_{III}$ has two layers of alternating spins in the [001] direction. GdN was found by our calculations to have a lower energy in the FM than in any of the AFM configurations, but GdP, GdAs, GdSb, and GdBi have $AFM_{II}$ as the lowest energy, followed by FM, then $AFM_{III}$ and $AFM_I$. The energies of the different configurations, given with respect to the FM state are given in Figure 8.

It is useful to analyze the energies using a Heisenberg Hamiltonian [41]

$$H = -2\sum_{i>j} J_{ij}\mathbf{S}_i \cdot \mathbf{S}_j. \quad (3)$$

We here adopt a quantum mechanical Heisenberg Hamiltonian with $S$=7/2, corresponding to the total localized magnetic moment for Gd where $L$=0. Alternatively, one could adopt a classical Heisenberg Hamiltonian with $S = \pm 1$ and this would simply renormalize the exchange interactions by a factor $S(S+1)$. Here we focus only on collinear magnetic arrangements, which are easily calculated by first principles to determine the exchange interactions, but, in principle, the same Hamiltonian can then be applied also to non-collinear and situations and to the statistical mechanics problem. Based on previous work,[34, 35] which found $J_3$ to be small, we adopt a model with only nearest, $J_1$, and second nearest-neighbor interactions, $J_2$. Positive coupling constants correspond to FM interactions. The energies for the four relevant magnetic configurations can be written as

$$E_{FM} = E_0 + S(S+1)(-12J_1 - 6J_2) \quad (4a)$$
$$E_{AFM_I} = E_0 + S(S+1)(4J_1 - 6J_2) \quad (4b)$$
$$E_{AFM_{II}} = E_0 + S(S+1)(6J_2) \quad (4c)$$
$$E_{AFM_{III}} = E_0 + S(S+1)(-4J_1 - 2J_2) \quad (4d)$$

with $S = 7/2$ or $S(S+1) = 63/4$. We extract $J_1$ and $J_2$ from the first-principles calculated energy differences $E(AFM_I) - E(FM)$ and $E(AFM_{II}) - E(FM)$ and use these values to test the accuracy for $E(AFM_{III}) - E(FM)$ . We emphasize here the $AFM_{II}$ configuration since this is the one actually found in experiment and in our calculations of GdP, GdAs, GdSb and GdBi to be the ground state. The results are shown in Table IV.

TABLE IV: Calculated and experimental[4] coupling constants $J_1$ and $J_2$ for Gd-pnictides, in K

| | $J_1$ | | $J_2$ | |
|---|---|---|---|---|
| (K) | Theory | Expt | Theory | Expt |
| GdN | 0.695 | 0.64 | $-0.322$ | 0.00 |
| GdP | 0.418 | 0.22 | $-0.498$ | $-0.34$ |
| GdAs | 0.363 | 0.08 | $-0.561$ | $-0.35$ |
| GdSb | 0.294 | 0.005 | $-0.721$ | $-0.60$ |
| GdBi | 0.253 | $-0.04$ | $-0.904$ | $-0.63$ |

TABLE V: Heisenberg model prediction for $AFM_{III}$-FM energy difference compared with direct calculation.

| (meV) | model | direct |
|---|---|---|
| GdN | 5.79 | 10.20 |
| GdP | 1.83 | 3.27 |
| GdAs | 0.93 | 2.74 |
| GdSb | $-0.71$ | 1.71 |
| GdBi | $-2.16$ | 0.35 |

First of all, we check the internal consistency of our model by comparing the predictions of the model for $AFM_{III}$-FM with the directly calculated results. This is shown in TableV. The model appears to generally underestimate the first-principles results, especially for GdN. However, for the rest of the series the error is less than 3 meV.

Next, we compare with the experimental values for the exchange parameters. The experimental estimates of $J_1$ and $J_2$ were obtained from the paramagnetic Curie-Weiss temperature, $\theta_P$, and critical field, $H_c$[4], using the mean field relations, given below. In principle, our procedure in which the parameters are directly given in terms of the energy differences is more direct since no mean field approximation is involved. Within the mean field approximation, one can also obtain the Néel temperature, although it is well known that mean field theory does not predict critical temperatures very accurately. The relevant equations are:

$$k_B T_N = -4S(S+1))J_2 \quad (5a)$$
$$k_B \theta_P = 4S(S+1)(2J_1 + J_2) \quad (5b)$$
$$g\mu_B H_c = -4S(6J_1 + 6J_2) \quad (5c)$$

We note that the calculated $J_2$ parameters are fairly close to the experimental values and follow the same trend, in particular, the values for P and As are very close but $|J_2|$ then increases significantly toward Sb and Bi. The $J_1$ interactions decrease from N to Bi but much more rapidly in the experimentally extracted values. For GdN, quite good agreement is obtained.

As a further way to analyze the comparison with experiment, we have calculated the directly observable quantities, $T_N$, $\theta_p$ and $H_c$ from our calculated parameters within mean field. This comparison is shown in Table VI. For GdN, which is ferromagnetic, the Curie temperature equals the paramagnetic Curie-Weiss expression given above within mean field theory. The average of the experimental values for $\theta_P$ (81 K) and $T_c$ (58 K), 69.5 K, is very close to the mean field value obtained from our calculated exchange parameters. For the other pnictides, our calculated value for $T_N$ overestimates the experimental values by about a factor 2, which is not unusual for mean field theory. It is encouraging though that the correct trend is obtained, note that $T_N$(calc)/2 has values of 17.5, 17.5, 22, and 27 K, respectively, very close to the data (15.9, 18.7, 23.4, and 25.8 K, respectively[29]). As for the paramagnetic Curie-Weiss temperature, we note that our values appear to differ from the experimental values by an almost constant shift of about 20 K, consistently being too high. In any case, they agree with the trend of an increasingly more negative $\theta_P$, a hallmark of antiferromagnetism. Finally, the critical field for aligning the magnetization with the external field below $T_N$ have the correct order of magnitude and underestimate the experimental values by 40 % to 9 %, the agreement becoming gradually better as we go to heavier pnictides. We did not calculate it for GdN since it is ferromagnetic. Considering that all of these values are based on very small energy differences of order meV, and that no adjustable parameters are used here that directly relate to the magnetic properties, we consider this quite good agreement with the experimental data.

Both our calculated values and the experimentally extracted exchange couplings indicate a decreasing trend of the exchange coupling parameters with the size of the anion. This is illustrated in Figure 9. The same trend was observed in Eu monochalcogenides, for $J_1$ but a weaker trend was obtained for their $J_2$, as pointed out by Kasuya and Li.[9]

TABLE VI: Comparison between mean-field predicted values and experimental[29] data for various magnetic properties.

| | $T_N$ (K) | | $\theta_P$ (K) | | $H_c$ (T) | |
|---|---|---|---|---|---|---|
| | Theory | Expt | Theory | Expt | Theory | Expt |
| GdN | 67.3[a] | 58.0 | 67.3 | 81.0 | – | 0.8 |
| GdP | 31.3 | 15.9 | 21.2 | 4.0 | 5.3 | 9.6 |
| GdAs | 35.3 | 18.7 | 10.4 | −11.8 | 13.0 | 16.7 |
| GdSb | 45.3 | 23.4 | −8.3 | −31.0 | 27.9 | 35.0 |
| GdBi | 56.9 | 25.8 | −25.0 | −45.0 | 42.7 | 42.0 |

[a] $T_c$

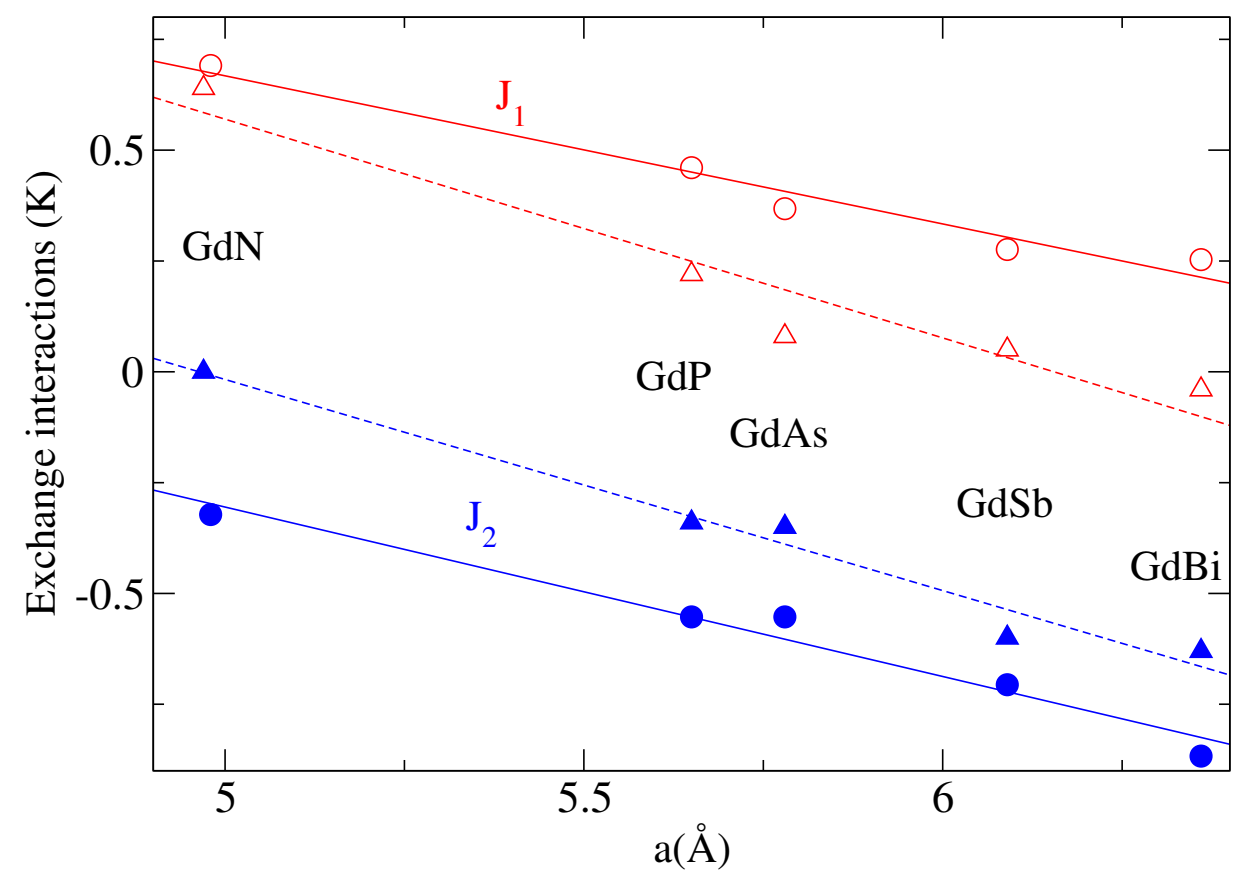

FIG. 9: Exchange interactions as function of lattice constant. Open symbols $J_1$, filled symbols $J_2$, circles: extracted from first principles calculated total energies, triangles, extracted from experimental data on $H_c$ and $\theta_p$ in mean field, [7]. The straight lines regressions are meant as guides for the eye.

## IV. SUMMARY

Band structure and total energy calculations were carried out for the entire series of Gd-pnictides. Both Hubbard-U terms for the $4f$ states and a Gd $5d$ band shift of empty states were included. The parameters were adjusted to some basic experimental such as the $4f$ occupied from unoccupied states splitting in photoemission and inverse photoemission and the optical absorption gap in GdN in the paramagnetic state. The large spin splitting of the majority and minority spin gaps (due to the spin inversion in valence and conduction band states) leads to a strong reduction of the gap from the paramagnetic to the ferromagnetic state. The almost negligibly small majority spin gap obtained for ferromagnetic GdN is consistent with the experimentally observed insulator to metal transition on going below the Curie temperature. The other pnictides were found to have an increasing downward shift of both occupied and unoccupied $4f$ states relative to the Fermi level, in qualitative agreement with photoemission data. They also indicate that the four heavier pnictides are semimetallic. Interestingly, we predict that GdP and GdAs could have half-metallic behavior in the ferromagnetic state, while this is not true for GdSb and GdBi. In any case, we expect a change in band overlap and hence carrier concentrations upon magnetizing these materials. The calculations also confirm the preference for a ferromagnetic ground state in GdN and antiferromagnetic ($AFM_{II}$) ground state for the other Gd-pnictides. First and second nearest neighbor Heisenberg exchange interactions were extracted from comparison of the energy differences of different magnetic configurations. They show an almost linear decreasing trend with lattice constant, in agreement with experimental data. The Curie temperature of GdN in mean field agrees well with the experimental data (to within about 10 K) and even better with the average of the experimental Curie temperature obtained from the peak in susceptibility and the Curie-Weiss temperature obtained from the high temperature behavior of the inverse susceptibility. The Néel temperatures for the other pnictides obtained within mean field theory form our calculated parameters overestimate the measured ones by about a factor 2 but follow the correct trend. The calculated Curie-Weiss temperatures appear to be shifted from the experimentally determined ones by a constant 20 K shift. The critical magnetic fields underestimate the measured ones by 10-40 % with increasingly better agreement for the heavier pnictides.

### Acknowledgments

This work was supported by the Office of Naval Research under grant number N00014-99-1-1073 and the National Science Foundation under grant number ECS-0223634.